\documentstyle[12pt]{article}
\input epsf

\newlength{\extraspace}
\newlength{\extraspaces}
\newcommand{\be}{\begin{equation}
\addtolength{\abovedisplayskip}{\extraspaces}
\addtolength{\belowdisplayskip}{\extraspaces}
\addtolength{\abovedisplayshortskip}{\extraspace}
\addtolength{\belowdisplayshortskip}{\extraspace}}
\newcommand{\ee}{\end{equation}}

\newcommand{\bq}{\begin{eqnarray}
\addtolength{\abovedisplayskip}{\extraspaces}
\addtolength{\belowdisplayskip}{\extraspaces}
\addtolength{\abovedisplayshortskip}{\extraspace}
\addtolength{\belowdisplayshortskip}{\extraspace}}
\newcommand{\eq}{\end{eqnarray}}

\newcommand{\newsection}[1]{
\vspace{15mm}
\pagebreak[3]
\addtocounter{section}{1}
\setcounter{equation}{0}
\setcounter{subsection}{0}
\setcounter{footnote}{0}
\begin{flushleft}
{\large\bf \thesection. #1}
\end{flushleft}
\nopagebreak
\medskip
\nopagebreak}

\begin{document}
\hbox{}
\nopagebreak
\vspace{-3cm}
\addtolength{\baselineskip}{.8mm}
\baselineskip=24pt
\begin{flushright}
{\sc OUTP}-97 P 21\\
hep-th@xxx/9705136\\
\today
\end{flushright}

\begin{center}
{\Large  A variational approach to the QCD wavefunctional:\\
 Calculation of the QCD $\beta$-function.}\\
\vspace{0.1in}
{\large William E. Brown and Ian I. Kogan}
\footnote{ On  leave of absence
from ITEP,
 B.Cheremyshkinskaya 25,  Moscow, 117259, Russia.}\\
{\it  Theoretical Physics, 1 Keble Road, \\
 Oxford, OX1 3NP \\
 U.K.} \\
\vspace{0.1in}

\vspace{0.1in}

{\sc  Abstract} 
\end{center}

\noindent
The $\beta$-function is calculated for an SU(N) Yang-Mills theory from
an ansatz for the vacuum wavefunctional.  Direct comparison is made
with the results of calculations of the $\beta$-function of QCD.  In
both cases the theories are asymptotically free.  The only difference
being in the numerical coefficient of the $\beta$-function, which is
found to be $-4$ from the ansatz and $-4+\frac{1}{3}$ from other QCD
calculations.  This is because, due to the constraint of Gauss' law
applied to the wavefunctional, transverse gluons (which contribute
the $\frac{1}{3}$) are omitted.  The renormalisation procedure is
understood in terms of `tadpole' and `horse-shoe' Feynman diagrams
which must be interpreted with a non-local propagator.   
\vfill

\newpage
\newsection{Introduction.}

\renewcommand{\footnotesize}{\small}

One of the main problems in modern quantum field theory is the
understanding of low energy phenomena in QCD, such as confinement,
chiral symmetry breaking, or, in more general terms, the strong
coupling problem.  To have analytic results for the ground state
of an asymptotically free non-abelian gauge theory, with the associated
enhanced understanding of the underlying physics, would be invaluable
in the understanding of these phenomena.  Although many promising
ideas have been suggested in the first nearly quarter century of QCD,
e.g. \cite{1}, we are still far from a completely satisfactory answer.

The arsenal of non-perturbative methods available to
tackle strongly interacting continuum theories is limited.  Although
in simple quantum mechanical problems a variational approach is often
easy to use - it is usually enough to know a few simple qualitative
features in order to set up a variational ansatz that will give good
results for the ground state energy and other vacuum expectation
values - this is not the case in quantum field theories, the
complexities of which pose difficult problems, as discussed by
Feynman, \cite{2}.

Recently, a variational approach based on a gauge invariant Gaussian
wavefunctional has been studied.  This method
was applied to QCD and QED$_3$, \cite{4} and \cite{3} respectively, where, although in its infancy, it has
independently verified many old, and given some new, results.  A
Gaussian approach to the wavefunctional of QCD (the so-called squeezed gluons) has been studied in
many different papers, see \cite{extra} and the references therein.  In the case of QED$_3$ the variational
method reproduced Polyakov's path integral results of the mass gap and
string tension of the theory, \cite{3}. 

The variational calculation carried out for the SU(N) purely Yang-Mills
theory in $3+1$ dimensions, \cite{4}, has found that the ground state energy is
minimal for a state which is different from the perturbative vacuum
even though the perturbative vacuum state was included in the
variational ansatz.  Dynamical scale generation takes place and the
gluon (SVZ) condensate in the best variational state was found to be
non-zero.  

The ansatz used for the vacuum wavefunctional was Gaussian but it was
also required to be gauge invariant.  To satisfy this, the Gaussian
wavefunctional is
projected onto the gauge invariant sector.  The form of the
wavefunctional is discussed briefly in the
next section and in detail in \cite{4}.  The variational parameter of
the theory is the mass
scale, $M$.  A non-zero value of $M$ for the minimal state
corresponds to a non-perturbative dynamical scale generation.

Within the variational calculation of \cite{4} it was conjectured that the
variational ansatz proposed yields a theory with a coupling constant
that runs as the coupling constant of (asymptotically free) QCD.  In
this paper we prove this conjecture.  It is shown
that the proposed variational ansatz yields an effective non-local,
non-linear sigma model in three dimensions which, when renormalised to
first order, has practically (a precise qualification of this word
will be given below) the same $\beta$-function as asymptotically free
QCD.  Further, the renormalisation procedure is interpreted in terms
of the Feynman diagrams included and it is found that, due to the
non-local nature of the propagator, a new `horse-shoe' diagram is
non-zero and makes a vital contribution.

In this paper, the $\beta$-function for the charge of the variational
ansatz is calculated to be
\begin{equation}
\beta(g)=-\frac{g^3}{(4\pi)^2}4C_2(G) +O(g^5)
\end{equation}
This should be directly compared with the known $\beta$ function for
QCD, \cite{8};
\begin{equation}
\beta(g)=-\frac{g^3}{(4\pi)^2}[(4-\frac{1}{3})C_2(G)-\frac{2}{3}n_fC(r)]+O(g^5)
\end{equation}
$n_f$ is the number of species of fermions in representation $r$,
which is zero for the purely Yang-Mills model.  For SU(N), $C_2(G)$,
the quadratic Casimir in the adjoint representation, is $N$.  We immediately see that the only
difference between the two $\beta$-functions is in the numerical terms
$4$ and $(4-\frac{1}{3})$.  The $4$ is due to the anti-screening
effect of longitudinal gluons.  The $\frac{1}{3}$ is due to the
screening effect of virtual transverse gluons.  Identically the same
$\beta$-function as QCD is not obtained because only longitudinal
gluons have been included within the ansatz.  The longitudinal gluons
are included because they satisfy Gauss' law with a source, which is
used as a constraint upon the vacuum wavefunctional in setting up the
variational ansatz.

In this paper the renormalisation calculation to first order is
presented and the renormalisation of the effective charge in the
variational wavefunctional is obtained.  In the first
section we shall describe the variational ansatz of \cite{4} and show
how this leads to an effective non-local, non-linear sigma model in
three dimensions.  In the second section we perform the
renormalisation group transformation, integrating over high momentum
dependent modes to yield an effective action for the low momentum
modes with a renormalised coupling constant, and we interpret the
renormalisation procedure by considering the Feynman diagrams which contribute.

\newsection{The variational ansatz.}

For a full discussion of the variational ansatz and all details of the
subsequent variational calculation the reader is directed to the
original paper, \cite{4}.  An overview of the variational ansatz and
the form of the effective action is
given in this section.

The SU(N) gauge theory is described by the Hamiltonian,
\begin{equation}
H= \int d^{3}x \left[{1\over 2}E^{a2}_i+{1\over 2}B^{a2}_i\right]
\label{ham}
\end{equation}
where
\begin{eqnarray}
E^a_i(x)&=&i{\delta\over \delta A^a_i(x)} \nonumber \\
B^a_i(x)&=&{1\over 2}\epsilon_{ijk}
\{\partial_jA_k^a(x)-\partial_kA^a_j(x)+gf^{abc}A_j^b(x)A_k^c(x)\}
\end{eqnarray}
and all physical states must satisfy the constraint of gauge 
invariance (Gauss' law);
\begin{equation}
G^a(x)\Psi[A]=\left[\partial_iE^a_i(x)-gf^{abc}A^b_i(x)E^c_i(x)\right]
\Psi[A]=0
\label{constr}
\end{equation}
Under a gauge transformation $U$ (generated by $G^a(x)$) the 
vector potential transforms as
\begin{equation}
A^a_i(x)\rightarrow  \ A^{Ua}_i(x)=S^{ab}(x)A_i^b(x)+\lambda_i^a(x)
\label{gt}
\end{equation}
where
\begin{eqnarray}
S^{ab}(x)={1\over 2}{\rm tr}\left(\tau^aU^\dagger\tau^bU\right); 
\ \ \ \lambda_{i}^a(x)={\frac{i}{g}{\rm tr}\left(\tau^aU^\dagger
\partial_{i}U\right)}
\label{defin}
\end{eqnarray}
and $\tau^a$ are traceless Hermitian N by N matrices satisfying 
$\rm{tr}(\tau^a\tau^b)=2\delta^{ab}$.  For SU(3) the algebra and
structure constants are defined, for example, in \cite{7}.

The initial ansatz for the ground state wavefunctional is of vital
importance.  It must incorporate the properties of all such physical
states and yet it must not lead to a solution which is incalculable 
if any progress is to be made.  In this formalism one calculates
expectation values of local operators with the ansatz for the ground
state, $\Psi$,
\begin{equation}
<O>=\frac{1}{Z}\int D\phi \Psi^*[\phi]O\Psi[\phi]
\end{equation}
and then minimises with respect to the variational parameter.  A
calculation of this kind is tantamount to evaluation 
of a Euclidean path integral with the square of the wavefunctional playing the 
role of the partition function. One should therefore be able to 
solve exactly a $d$-dimensional field theory with the action 
\begin{equation}
S[\phi]=-{\rm log}\Psi^*[\phi]\Psi[\phi]
\label{act1}
\end{equation}
Since in dimension $d>1$ the only theories one can solve exactly 
are free field theories, the requirement of calculability almost 
unavoidably restricts the possible form of the wavefunctional to a Gaussian 
(or, as it is sometimes called, squeezed) state:
\begin{eqnarray}
\Psi[A_i^a]=\exp\left\{ - \frac{1}{2}\int d^{3}x d^{3}y  
 \left[A_i^a(x)-\zeta_i^a(x)\right]
 (G^{-1})^{ab}_{ij}(x,y)
 \left[A_j^b(y)-\zeta_j^b(y)\right]\right\}
\end{eqnarray}
with $\zeta(x)$ and $G(x,y)$  being c-number functions.  The
requirement of translational invariance usually gives 
further restrictions: $\zeta(x)=\rm const$,~ $G(x,y)=G(x-y)$.

There is, however, one obvious difficulty with this idea.  It is
impossible to write down a Gaussian wavefunctional which satisfies the
constraint of gauge invariance.  Under the gauge transformation the
wavefunctional transforms as
\begin{equation}
\Psi[A_i^a]\rightarrow\Psi[(A^{U})_i^{a}]
\end{equation}
In the abelian case it is enough to take $\partial_iG^{-1}_{ij}=0$ 
to satisfy the constraint of gauge invariance.  In the non-abelian 
case, however, due to the homogeneous piece in the gauge transformation 
(\ref{gt}), no gauge invariant Gaussian wavefunctional exists.

The proposed solution to this problem was to simply project the
Gaussian wavefunctional onto the gauge invariant sector and to
restrict the calculation to the case of zero classical fields
($\zeta=0$). The variational ansatz is therefore
\begin{equation}
\Psi[A_i^a]=\int DU(x)
 \exp\left\{-{1\over 2}\int d^{3}x d^{3}y 
\ A_i^{Ua}(x)G^{-1ab}_{ij}(x-y)\ A_j^{Ub}(y)\right\}
\label{an}
\end{equation}
with $A_i^{Ua}$ defined in (\ref{gt}) and the integration 
performed over the space of special unitary matrices with 
the $SU(N)$ group invariant measure.

Further restrictions upon the form of $G$ lead to considerable
simplifications.  Firstly, only matrices of the form 
\begin{equation}
G^{ab}_{ij}(x-y)=\delta^{ab}\delta_{ij}G(x-y)
\label{an1}
\end{equation}
are considered.  This is certainly the correct form in the
perturbative regime.  If it was not for the integration over the
group, $G_{ij}^{-1}$ would be precisely the (equal time) propagator of
the electric field.  Due to the integration over the group, however,
the actual propagator is the transverse part of $G^{-1}$.  The
longitudinal part $\partial_i G_{ij}^{-1}$ drops out of all physical
quantities, giving, without any loss of
generality at the perturbative
level, $G_{ij} \sim \delta_{ij}$.  Also, in the leading order in
perturbation theory, the non-abelian character of the gauge group is
not important.  The $\delta^{ab}$ structure is then obvious.

The form of $G$ can be restricted further using additional
perturbative information.  The theory is asymptotically free.  This
means that the short distance asymptotics of correlation functions
must be the same as in the perturbation theory.  Since $G^{-1}$ is
directly related to correlation functions of gauge invariant
quantities in perturbation theory, it is taken to have the form,
\begin{equation}
G^{-1}(x)\rightarrow  {1\over x^4}, ~~~~~ x \rightarrow 0
\label{an2}
\end{equation}
The non-perturbative theory is also expected to have a gap. 
In other words, the correlation functions should decay to zero 
at some distance scale,
\begin{equation}
G(x)\sim 0, \ \ x>{1\over M}
\label{an3}
\end{equation}
The variational ansatz is built in the simplest 
possible way.  $M$ is taken to be the only variational 
parameter and this is done by choosing $G(x)$ to be of a form that has
the ultra-violet and infra-red asymptotics of (\ref{an2}) and
(\ref{an3}).  A non-zero result for $M$ means a non-perturbative
dynamical scale generation in the Yang-Mills vacuum.  The form of
$G^{-1}$ used in this paper and the variational
calculation of \cite{4} has the Fourier transform
\begin{eqnarray}
G^{-1}(k) = \left\{ \begin{array}{ll} \sqrt{ k^{2} ~} & 
\mbox{ if  $ k^2>M^2$}\\
 M &  \mbox{ if $k^2<M^2$} 
\end{array} 
\right.
\label{an4}
\end{eqnarray}
Equation (\ref{an}) together with equations (\ref{an1}) and
(\ref{an4}) define our variational ansatz.  

It will now be shown that the action (\ref{act1}) is in fact a
non-local, non-linear sigma model.  Again, the reader is referred to
\cite{4} for the original account.  With the given ansatz, the
remaining problem is to calculate expectation values of local
operators, such as,
\begin{equation}
<O>= {1\over Z}\int DU DU' <O>_{A}
\end{equation}
where
\begin{eqnarray}
 <O>_{A}&=& \int DA e^{-{1\over 2}
\int dx dy  A_i^{Ua}(x)G^{-1}(x-y)A_i^{Ua}(y)}~ O~  
e^{-{1\over 2}\int dx' dy' A_j^{U'b}(x')G^{-1}(x'-y')A_j^{U'b}(y')}\nonumber \\
&=& \int DA e^{-{1\over 2}\int dx dy A_i^{Ua}(x)
G^{-1}(x-y) A_i^{Ua}(y) }O e^{-{1\over 2}\int dx' dy' 
A_j^b(x')G^{-1}(x'-y') A_j^b(y')} 
\end{eqnarray}
where, since only gauge invariant operators are to be considered, the
change of variable $A_i \to A_i^{-U'}$ has rendered one of the group
integrations redundant.

For convenience, the definition
\begin{equation}
a_i^a(x)=\int d^3y d^3z \lambda_i^b(y) G^{-1}(y-z) S^{bc}(z)({\cal
M}^{-1})^{ca}(z,x)
\end{equation}
is made so that the Gaussian integration over $A_i$ is $\int DA \exp
[-\frac{1}{2}(A+a){\cal M}(A+a)]$.  Changing variables and performing
the integration yields the following form of the
normalization factor $Z$,
\begin{equation}
Z=\int DU\exp\{-\Gamma[U]\}
\label{sigma}
\end{equation}
with an effective action 
\begin{equation}
\Gamma[U]={1\over 2} \rm{Tr} \ln{\cal M}
+{1\over 2}\lambda^a \Delta^{ac}\lambda^c
\label{action}
\end{equation}
where
\begin{equation}
\Delta^{ac}(x,y)=[G(x-y)\delta^{ac}+S^{ab}(x)G(x-y)S^{Tbc}(y)]^{-1}
\label{invop}
\end{equation}
is the `effective inverse propagator' and multiplication is understood
as the matrix multiplication with indices: colour $a$, space $i$ and position 
(the values of space  coordinates) $x$,
i.e. 
\begin{equation}
(A B)_{ik}^{ac}(x,z) = \int d^{3}y A_{ij}^{ab}(x,y)
 B_{jk}^{bc}(y,z),~~~~~~~~~~ 
 \lambda O \lambda = \int d^3x d^3y \lambda^{a}_{i}(x) O^{ij}_{ab}(x-y)
 \lambda^{b}_{j}(y)
\end{equation}
 The trace $ \rm{Tr} $ is understood as a trace over all three types
of indices.  In equation (\ref{action}) we have defined
\begin{equation}
S^{ab}_{ij}(x,y)=S^{ab}(x)\delta_{ij}\delta(x-y), \ \ 
{\cal  M}^{ab}_{ij}(x,y)=
[S^{Tac}(x)S^{cb}(y)+\delta^{ab}]G^{-1}(x-y)\delta_{ij}
\label{def1}
\end{equation}
where $
S^{ab}(x)={1\over 2}\rm{tr}\left(\tau^aU^\dagger\tau^bU\right)$ and 
$~\lambda_i^a(x)={i\over g}{\rm tr}\left(\tau^aU^\dagger
\partial_{i} U\right)$ were defined in (\ref{defin}) and $\rm{tr}$ is a
trace over colour indices only.  One should also note here another
useful definition, the completeness condition for SU(N);
\begin{equation}
\tau_{ij}^a \tau_{kl}^a = 2(\delta_{il} \delta_{jk}
-\frac{1}{N}\delta_{ij} \delta_{kl})
\label{comp}
\end{equation}

The path integral (\ref{sigma}) defines a partition function of a
non-linear sigma model in three dimensional Euclidean space.  The
action of this sigma model is rather complicated.  It is a non-local
and a non-polynomial functional of $U(x)$.  We shall see how the
coupling appears in the effective action in both high and low momentum
cases.  Various approximations are made here which are
explored much more rigorously in the calculation of the succeeding section.

For high momentum modes, with the standard parametrization
$U(x)= \exp[i\frac{g}{2} \phi^a(x) \tau^a]$, one gets
$\lambda_i^a(x)=-\partial_i \phi^a (x) +O(g)$,
$S^{ab}(x)=\delta^{ab}+O(g)$ and the leading order term in the action
becomes:
\begin{equation}
\frac{1}{4} \int d^3x d^3y \partial_i \phi^a(x) G^{-1}(x-y) \partial_i
\phi^a(y)
\end{equation}
This is just a free theory with a non-standard propagator;
\begin{equation}
<\phi^a(x) \phi^b(y)> = 2\delta^{ab} [\partial_i^x \partial_i^y G^{-1}
(x-y)]^{-1}= 2\delta^{ab} \int \frac{d^3k}{(2\pi)^3}\frac{\exp[ik.(x-y)]}{k^3}
\end{equation}

For the low momentum modes, to a first level of approximation, the
space dependence of $S^{ab}(x)$ is ignored in the term $SGS^T$ giving,
with the fact that $S$ is an orthogonal matrix, the approximation
$SGS^T\rightarrow G$.  Then, using the completeness condition,
(\ref{comp}), and the fact that ${\rm tr}(U^\dagger \partial_i U)=0$ we can
write
\begin{equation}
\lambda^{a}_{i}(x)\lambda^{a}_{i}(x) = -(1/g^2) 
{\rm tr}\left(\tau^aU^\dagger\partial_iU\right)
{\rm tr}\left(\tau^aU^\dagger\partial_iU\right)
= -(2/g^{2}) {\rm tr}\left(U^\dagger\partial_iU \ 
U^\dagger\partial_iU\right)
\end{equation}
In this approximation the action becomes
\begin{equation}
\frac{1}{2g^2}{\rm tr} \int d^3x d^3y \partial_i U^\dagger (x) G^{-1}(x-y)
\partial_i U(y) =\frac{M}{2g^2}{\rm tr} \int d^3x \partial_i U^\dagger (x)
\partial_i U(x)+...
\end{equation}
where $+...$ corresponds to all the higher terms in $g$.

\newsection{Calculation of the $\beta $-function.}

Having obtained the effective action of (\ref{action}), which is
a non-local, non-linear sigma model in three dimensions, we shall
proceed to calculate the $\beta$ function.  The form of the
$\beta$-function is deduced from the renormalised coupling constant,
in the spirit of \cite{6} where a similar calculation was performed for
the non-linear sigma model in two dimensions.  We shall perform the
renormalisation group transformation by integrating over high momentum
dependent modes (containing Fourier components $k>M$) leaving an effective
action for low momentum dependent modes (containing Fourier components
$k<M$).  The coupling constant of the effective action is renormalised
to first order; up to terms quadratic in the high momentum modes.  A
physical interpretation of these terms by considering the
corresponding Feynman diagrams is given in section 4.  The high
momentum modes are the quantum field, and the low momentum modes are
the classical field, of the background field method.

Next we shall discuss the decomposition of the group
elements into high and low momentum dependent modes and in the second
subsection we shall explicitly calculate the $\beta$-function.

\subsection{Quadratic approximation for the high momentum modes.}

The ansatz proposed for the decomposition of group elements into high
and low momentum dependent modes is
\begin{equation}
U(x)=U_L(x) U_H(x)
\label{dec}
\end{equation}
where $U_L(x)$ contains Fourier components $k<M$ and $U_H(x)$ contains
Fourier components $k>M$.  This can be considered as a decomposition
of the group parameter, $\phi^a(x)$.  If we write
$U(x)=\exp[\frac{ig}{2}\phi^a(x)\tau^a]$ then the decomposition can be
written as $\phi^a(x) = \phi_H^a(x) + \phi_L^a(x)$, where
$\phi_{H,L}^a(x)$ are the group parameters of $U_{H,L}(x)$,
respectively, and $\phi_H^a$ and $\phi_L^a$ are taken to be
orthogonal; $\phi_H^a(x) \phi_L^a(x) = 0$.  

Written in terms of the group parameters, we can explicitly see that
this ansatz has similarities to that used by Polyakov in his treatment of
the non-linear sigma model in two dimensions, \cite{6}, but without
the normalisation of
$\phi_L^a(x)$ constrained to be that of $\phi^a(x)$.  In his ansatz,
Polyakov forced $|\phi^a|^2=|\phi_L^a|^2$, proposing
\begin{equation}
\phi^a(x)=\phi_L^a(x)(1-|\phi_H^a|^2)^{\frac{1}{2}} +
\phi_{i,H}(x)e_i^a(x)
\end{equation}
which has been written in the notation of this paper.  $e_i^a(x)$ form
a complete basis of unit vectors orthonormal to $\phi_L^a(x)$.  We do
not use this ansatz for the decomposition, however, because for each
order of $g$ in the calculation it mixes the high and low momentum
group parameters making the desired decomposition (\ref{dec})
difficult.

For the rest of this section we shall employ the decomposition
(\ref{dec}) to write the effective Lagrangian for $\lambda_{i,L}^a$ up
to terms of $O(g^2)$.  This corresponds to all terms up to, and
including, those quadratic in the field $\phi$.  

It is necessary to first note two identities for $\lambda_i^a$ and
$S^{ab}$.  Using the completenes condition for SU(N), (\ref{comp}),
with the cyclic property of the trace and the fact that $\rm{tr}[U\tau^a
U^{+}]=\rm{tr}[\tau^a]=0$ we find,
\begin{equation}
S^{ac}(x)=S_H^{ab}(x) S_L^{bc}(x)
\label{Sdec}
\end{equation}
and
\begin{equation}
\lambda_i^a(x)=S_H^{ab} \lambda_{i,L}^b(x) + \lambda_{i,H}^a(x)
\label{ldec}
\end{equation}

Using the same mathematical properties, we should also note that
$S^{ab}$ is an orthogonal matrix,
\begin{equation}
S^{ab}(x) S^{Tbc}(x) = S^{ab}(x) S^{cb}(x) = \frac{1}{4}\tau_{ij}^b
\tau_{kl}^b (U(x) \tau^a U^{\dagger}(x))_{ij} (U(x) \tau^c
U^{\dagger}(x))_{kl} = \frac{1}{2}tr[\tau^a \tau^c] = \delta^{ac}
\end{equation}

First, we shall evaluate the inverse effective propagator,
$\Delta^{ac}(x,y)$ (\ref{invop}), up to terms quadratic in the
coupling constant.  To do this we write,
\begin{equation}
U_H(x)=\exp[\frac{ig}{2}\phi^a(x) \tau^a]=1+\frac{ig}{2}\phi^a(x)
\tau^a -\frac{g^2}{8}(\phi^a(x) \tau^a)^2 + O(g^3)
\end{equation}
We shall use the following normalisation of the generators of SU(N),
(e.g.  with constants defined for SU(3) in \cite{7}),
\begin{eqnarray}
[\tau^a,\tau^b]&=&2if^{abc}\tau^c    \\  
\frac{1}{2}tr[\tau^a\tau^b]&=&\delta^{ab} 
\end{eqnarray}
We find,
\begin{eqnarray}
S_H^{ad}(x) &=& \frac{1}{2} tr[\tau^a, \tau^d]
-\frac{ig}{4}\phi^b(x)tr[\tau^a[\tau^b,\tau^d]] \\ \nonumber
& & - \frac{g^2}{16}\phi^b(x)\phi^c(x)tr[\tau^a[\tau^b,[\tau^c,\tau^d]]]
+O(g^3) \\ \nonumber
&=&\delta^{ad}-gf^{adb}\phi^b(x)-\frac{g^2}{2}f^{abe}f^{dce}\phi^b(x)\phi^c(x)
+O(g^3)
\label{S}
\end{eqnarray}
Therefore, using (\ref{Sdec}) and considering the low momentum group elements, $U_L(x)$, to
be slowly varying, such that $U_L^{\dagger}(x) U_L(y) \simeq 1$ and
$S_L^{db}(x) S_L^{eb}(y) \simeq \delta^{de}$, we find,
\begin{eqnarray}
S^{ab}(x) S^{Tbc}(y) &\simeq& S_H^{ad}(x) S_H^{cd}(y) \\ \nonumber
&=& \delta^{ac} - g\Delta_1^{ac}(x,y) - \frac{g^2}{2}
\Delta_2^{ac}(x,y) + O(g^3)
\end{eqnarray}
where
\begin{eqnarray}
\Delta_1^{ac}(x,y) &=& f^{acb}(\phi^b(x)-\phi^b(y)) \\
\Delta_2^{ac}(x,y) &=& f^{abe}f^{cge}(\phi^b(x)\phi^g(x)+\phi^b(y)\phi^g(y)-  
2\phi^b(x)\phi^g(y))
\end{eqnarray}
We now re-write $\Delta^{ac}(x,y)$ as,
\begin{eqnarray}
\Delta^{ac}(x,y) &=& G^{-1}(x-y) [2\delta^{ac}]^{-1}[1-
g\frac{[2\delta^{ac}]^{-1}}{\rm{tr}[\delta^{aa}]} \Delta_1^{ac}(x,y) \\
\nonumber
& & -\frac{g^2}{2} \frac{[2\delta^{ac}]^{-1}}{\rm{tr}[\delta^{aa}]}
\Delta_2^{ac}(x,y) +O(g^3)]^{-1} \\ \nonumber
&=&\frac{\delta^{ac}}{2} G^{-1}(x-y)[1+ \frac{g^2}{4\rm{tr}[\delta^{aa}]}
\Delta_2^{aa}(x,y)] + O(g^3)
\end{eqnarray}
where $[\delta^{ac}]^{-1}=\delta^{ac}$, $\delta^{ac}
[\delta^{ac}]^{-1} = \rm{tr}[\delta^{aa}]$.  Also,
\begin{eqnarray}
\Delta_1^{aa}(x,y) &=& 0 \\
\Delta_2^{aa}(x,y) &=& C_2(G) (\phi^b(x)\phi^b(x) + \phi^b(y)\phi^b(y)
- 2\phi^b(x)\phi^b(y))
\label{delta}
\end{eqnarray}
$C_2(G)$ is the second Casimir operator and $G$ denotes the adjoint
representation in this case.  For SU(N),
$C_2(G)=N$.  Equation (\ref{delta}) is obtained using the relation,
\begin{equation}
f^{acd}f^{bcd}=C_2(G)\delta^{ab}
\end{equation}

Next we need to evaluate $\lambda_i^a(x)$ up to $O(g^2)$.  Using
(\ref{ldec}), (3.9) and following similar intermediary steps to
the calculation of (3.9), we find,
\begin{eqnarray}
\lambda_{i,H}^a(x) &=& -\partial_i \phi^a(x) +O(g^3) \\
\lambda_i^a(x) &=& \lambda_{i,L}^a(x)-\partial_i \phi^a(x) -
gf^{abc}\lambda_{i,L}^b(x) \phi^c(x) \\ \nonumber
& & - \frac{g^2}{2} f^{ace}f^{bde}
\lambda_{i,L}^b(x) \phi^c(x) \phi^d(x) +O(g^3)
\end{eqnarray}

Therefore, we can now write the effective action as,
\begin{equation}
\Gamma[U]=\int d^3x d^3y \Gamma[x,y]
\end{equation}
where
\begin{eqnarray}
\Gamma(x,y) &=& \frac{1}{4} \partial_i \phi^a(x)G^{-1}(x-y)\partial_i
\phi^a(y) \\ \nonumber
& & +\frac{1}{4}\lambda_{i,L}^a(x) G^{-1}(x-y)
\lambda_{i,L}^a(y)\\ \nonumber
& & + \frac{g}{4}f^{abg}[\lambda_{i,L}^b(x)\phi^g(x)G^{-1}(x-y)
\partial_i \phi^a(y) \\ \nonumber
& & + \partial_i \phi^a(x)G^{-1}(x-y)
\lambda_{i,L}^b(y)\phi^g(y)] \\ \nonumber
& & - \frac{g^2}{8}f^{age}f^{bde} \lambda_{i,L}^b(x)G^{-1}(x-y)
\lambda_{i,L}^a(y) [\phi^g(x)\phi^d(x) \\ \nonumber
& & +\phi^g(y)\phi^d(y) -
2\phi^g(x)\phi^d(y)] \\ \nonumber
& & + \frac{g^2}{16} \lambda_{i,L}^a(x) G^{-1}(x-y) \lambda_{i,L}^a(y)
\frac{C_2(G)}{tr[\delta^{aa}]} [\phi^b(x)\phi^b(x) \\ \nonumber
& & +\phi^b(y)\phi^b(y)
-2\phi^b(x)\phi^b(y)]
\label{eff}
\end{eqnarray}
where $\Gamma(x,y)$ is the effective Lagrangian.

\subsection{Renormalisation group transformation.}

The renormalisation group transformation is performed by integrating
over the high momentum dependent field, $\phi^a(x)$.  This is akin to
integrating over the fluctuating (quantum) fields in the background
field method.

We consider
\begin{equation}
\Gamma^0(x,y)=\frac{1}{4} \partial_i \phi^a(x) G^{-1}(x-y) \partial_i
\phi^a(y)
\label{eff0}
\end{equation}
to be the zeroth order Lagrangian for the field $\phi^a(x)$.  All
other terms involving $\phi^a$ are treated as perturbations of this
Lagrangian.  (\ref{eff0}) implies the two-point Green's function
\begin{equation}
<\phi^a(x)\phi^b(y)>=2\delta^{ab}[\partial_i^x \partial_i^y
G^{-1}(x-y)]^{-1}
\end{equation}
$G^{-1}(x-y)$ is defined in (\ref{an4}).  So, because $\phi^a(x)$ is
defined to have Fourier components $k>M$, we can write,
\begin{equation}
<\phi^a(x)\phi^b(y)>=2\delta^{ab} \int \frac{d^3k}{(2\pi)^3}
\frac{\exp[ik.(x-y)]}{k^3} 
\label{Four}
\end{equation}
with the integral performed over the limits $M<k<\Lambda$,
$0<\phi<2\pi$, $0<\theta<\pi$, where $\Lambda$ is the ultra-violet
cut-off.  

Treating terms other than $\Gamma^0$ as perturbations, we see that
\begin{equation}
\int D \phi\exp[\int d^3x d^3y \{ \Gamma^0(x,y) + F[\phi] \}] =
\exp[\int d^3x d^3y<F[\phi ]>]
\end{equation} 
So we now write the effective Lagrangian as
\begin{eqnarray}
\Gamma_L(x,y) &=& \frac{1}{4}\lambda_{i,L}^a(x) G^{-1}(x-y)
\lambda_{i,L}^a(y)\\ \nonumber
& & + \frac{g}{4}f^{abg}[\lambda_{i,L}^b(x) G^{-1}(x-y)<\phi^g(x)
\partial_i \phi^a(y)> \\ \nonumber
& & + <\partial_i \phi^a(x)\phi^g(y)>  G^{-1}(x-y)
\lambda_{i,L}^b(y)] \\ \nonumber
& & - \frac{g^2}{8}f^{age}f^{bde} \lambda_{i,L}^b(x)G^{-1}(x-y)
\lambda_{i,L}^a(y) [<\phi^g(x)\phi^d(x)> \\ \nonumber
& & +<\phi^g(y)\phi^d(y)> -
2<\phi^g(x)\phi^d(y)>] \\ \nonumber
& & + \frac{g^2}{16} \lambda_{i,L}^a(x) G^{-1}(x-y) \lambda_{i,L}^a(y)
\frac{C_2(G)}{tr[\delta^{aa}]} [<\phi^b(x)\phi^b(x)> \\ \nonumber
& & +<\phi^b(y)\phi^b(y)>
-2<\phi^b(x)\phi^b(y)>]
\label{effx}
\end{eqnarray}

First considering the terms of $O(g)$ we see that both their
contributions are zero.  This is explicitly seen as both of the
correlations $<\phi^g(x) \partial_i \phi^a(y)>$ and $<\partial_i
\phi^a(x)\phi^g(y)>$ have the structure $\delta^{ag}$ preceeded by the
totally antisymmetric group structure constant $f^{abg}$.

Considering the terms of $O(g^2)$, we need to examine the evaluation
of (\ref{Four}) in some detail.  First, we note that,
\begin{equation}
<\phi^a(x) \phi^b(x)>=<\phi^a(y)
\phi^b(y)>= \frac{\delta^{ab}}{\pi^2}\log \frac{\Lambda}{M}
\label{ab1}
\end{equation}
which occurs in (\ref{effx}) in terms such as
\begin{equation}
\frac{g^2}{16} \lambda_{i,L}^a(x) G^{-1}(x-y) \lambda_{i,L}^a(y)
\frac{C_2(G)}{{\rm tr}[\delta^{aa}]} <\phi^b(x)\phi^b(x)>
\end{equation}
This can be represented by a Feynman diagram, Fig 1, which shows a
tadpole diagram.
\begin{figure}
\epsfbox{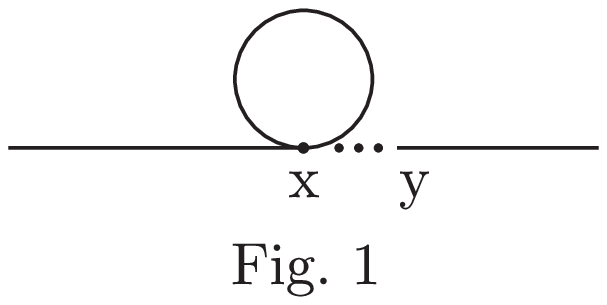}
\label{f1}
\end{figure}
The external lines correspond to low momentum fields, $U(x)$, and the internal
loop represents the integration over high momentum fields.  The dotted line corresponds to $G^{-1}(x-y)$.  It is important to note that in the
region corresponding to $k<M$, $G^{-1}(x-y)=M\delta(x-y)$ and the
propagator becomes local, associating either end of the dotted line,
and the standard tadpole diagram is recovered.  The evaluation of
(\ref{ab1}) is, however, the same at all scales of spatial separation, $|x-y|$.

To evaluate the other terms in $\Gamma_L (x,y)$, we write (\ref{Four})
in the form,
\begin{equation}
<\phi^a(x) \phi^b(y)> = \frac{\delta^{ab}}{\pi^2}
\int_{M|x-y|}^{\Lambda|x-y|} dt \frac{\sin t}{t^2}
\label{ab2}
\end{equation}
where the change of variable $t=k|x-y|$ has been made.  $t$ is a
dimensionless variable.  We shall introduce a scale, $\mu$, such that
for $t<\mu$, $\sin t \simeq t$.  This allows us to write,
\begin{equation}
<\phi^a(x) \phi^b(y)> = \left\{ \begin{array}{cc} a & \mbox{ $M|x-y|>\mu$} \\
-\frac{\delta^{ab}}{\pi^2} \log [M|x-y|] +b & \mbox{ $M|x-y|<\mu$} 
\end{array}
\right.
\label{ab3}
\end{equation}
where $a$ and $b$ are finite contributions which are independent of
$M$ and $\Lambda$ and hence are ignored in the following.  All $\mu$
dependence is in the finite terms.

The correlations $<\phi^a(x) \phi^b(y)>$  appear in $\Gamma_L (x,y)$ in
terms such as,
\begin{equation}
-\frac{g^2}{8} \lambda_{i,L}^a(x) G^{-1}(x-y) \lambda_{i,L}^a(y)
\frac{C_2(G)}{{\rm tr}[\delta^{aa}]} <\phi^b(x)\phi^b(y)>
\label{d1}
\end{equation}
which can also be interpreted in terms of Feynman diagrams.  Fig. 2
shows the diagram corresponding to (\ref{d1}).  It depicts a
horse-shoe.  As in Fig. 1, the external lines correspond to the low
momentum fields, $U_L(x)$, the horse-shoe represents the integration
over the high momentum fields and the dotted line again represents
$G^{-1}(x-y)$.  
\begin{figure}
\epsfbox{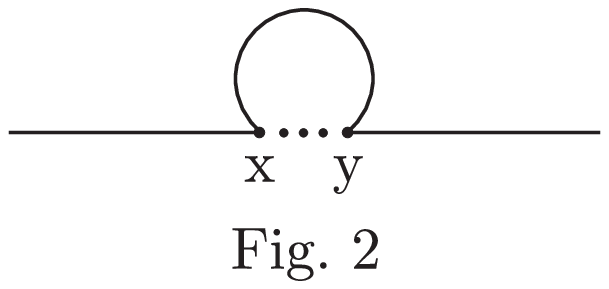}
\label{f2}
\end{figure}
The horse-shoe diagram is unimportant in the region $\frac{\mu}{M}<|x-y|$ as it
only gives a finite contribution, independent of a cut-off or $M$.  In
the region $0<|x-y|<\frac{\mu}{M}$ the horse-shoe diagram is important and
contributes $\frac{\delta^{ab}}{\pi^2}\log \frac{1}{M|x-y|}$,
(\ref{ab3}).  We must
not here immediately interpret $|x-y|$ to therefore be the inverse of
the cut-off but rather take our interpretation from the
form of the resulting Lagrangian.

Substitution of (\ref{ab1}) and (\ref{ab3}) into $\Gamma_L(x,y)$ yields
the effective action for the low momentum modes with a renormalised coupling
constant, $\tilde{g}$.  With
$\lambda_{i,L}^a=\frac{i}{\tilde{g}}{\rm tr}[\tau^a U_L^{+} \partial_i U_L]$
this can now be written as,
\begin{eqnarray}
\Gamma_L(x,y) &=& \frac{1}{4}\lambda_{i,L}^a(x)G^{-1}(x-y)\lambda_{i,L}^a(y)
\\ \nonumber
&=& \frac{1}{2\tilde{g}^2} {\rm tr}[\partial_i U_L(x) G^{-1}(x-y) \partial_i U_L^{\dagger}(y)]
\label{renact}
\end{eqnarray}
where 
\begin{eqnarray}
\tilde{g}^2 = \left\{ \begin{array}{cc} g^2(1+\frac{g^2}{2\pi^2}
C_2(G) \log \frac{\Lambda}{M}) & \mbox{ $M|x-y|>\mu$} \\ 
g^2(1+\frac{g^2}{2\pi^2} C_2(G) \log \Lambda|x-y|) & \mbox{ $M|x-y|<\mu$}
\end{array}
\right.
\label{gt2}
\end{eqnarray}

$\beta(g)$ is calculated using the standard definition,
\begin{equation}
\beta(g)=M\frac{\partial}{\partial M} \tilde{g}|_{g,\Lambda}
\end{equation}
but first we need to interpret the appearance of $|x-y|$ in
(\ref{gt2}).  For $|x-y|> \frac{\mu}{M}$, $M$ is the renormalisation
scale, whereas, for $|x-y|< \frac{\mu}{M}$, $\frac{1}{|x-y|}$ should be
interpreted as the renormalisation scale.  Therefore, we can re-write
$\tilde{g}$ as,
\begin{equation}
\tilde{g} = g + \frac{g^3}{(4\pi)^2} 4C_2(G) \log \frac{\Lambda}{M'}
+O(g^5)
\end{equation}
where
\begin{eqnarray}
M' = \left\{ \begin{array}{cc} M & \mbox{ $|x-y|>\frac{\mu}{M}$} \\
\frac{1}{|x-y|} & \mbox{ $|x-y|<\frac{\mu}{M}$}
\end{array}
\right.
\label{xx}
\end{eqnarray}

Thus we obtain,
\begin{equation}
\beta(g) = M' \frac{\partial}{\partial M'} \tilde{g}|_{g,\Lambda} =
-\frac{g^3}{(4\pi)^2} 4C_2(G) + O(g^5)
\end{equation}
This should be compared to the standard $\beta$-function for QCD,
e.g. \cite{8},
\begin{equation}
\beta(g) = -\frac{g^3}{(4\pi)^2}[(4-\frac{1}{3})C_2(G) -\frac{2}{3}n_f]+O(g^5)
\end{equation}
In this paper and in \cite{4}, a variational approach to QCD has been
considered in the absence of fermions, $n_f=0$.  With a gauge group SU(N), $C_2(G)=N$.
The discrepancy between $4$ in the result of this paper and the
$\frac{11}{3}$ of the standard QCD $\beta$-function quoted above is
due to the fact that only longitudinal gluons are included in the
variational ansatz and that transverse gluons, which would contribute
the extra $\frac{1}{3}$, are omitted.  This is as expected because in
the creation of the ansatz it was required that the wavefunctional
obey Gauss' law, which is indeed satisfied by longitudinal photons
with a source.

The  representation of a $\beta$-function coefficient as the sum of two
  contributions  proportional to $-4$ and $1/3$ is not new at all.
 It has been known for a long time, \cite{no} and \cite{h} that in the
background field method   the
contribution of charged  particles with spin $S$ to the
one-loop $\beta$-function is given by   
\begin{equation}
\beta_{S}(g) = - g^3\frac{(-1)^{2S}}{(4\pi)^2}
[(2S)^2-\frac{1}{3})
\end{equation}
 where we omit the group factor. The  asymptotically free (for integer
 spins) ``spin''
factor $4S^2$, where $S=1$ for the vector field $A$, gives $4$ which
is precisely what we have obtained in our calculations.
 As was discussed in \cite{h}, this spin factor  is related to
 the influence of the background field on the electric  dipole
 moment density, which includes contributions from the time-like
 polarization states of the massless vector fields in the case of the Feynman 
 gauge.  In a Coulomb gauge, where the time-like modes do not exist
 as dynamical degrees of freedom,
  one can see that the electric dipole density appears because of
 Gauss' law. This  explains why in our method, where  the Gauss' law
 is  implemented by construction, we  obtained the same result.
 
It is interesting to note that the same decomposition $4 - 1/3$
  comes from the results for the
calculation of the pre-exponential factor, or the renormalisation of
the charge, for the BPST instanton within the path integral formalism
of QCD, \cite{9}.  In this work, the vector field is split into
components, $A_{\mu}^a=A_{\mu}^{a(inst)} + a_{\mu}^a$, and the action
expanded in terms of the deviation $a_{\mu}^a$ from the instanton
field $A_{\mu}^{a(inst)}$.  Analysis of the resulting path integral
and examination of the contribution of the zero-frequency modes yields,
\begin{equation}
\frac{<0|0_T>_{ins}^{Reg}}{<0|0_T>_{p.th.}}=const. \int \frac{d^4x
d\rho}{\rho^5} S_0^4 \exp [-S_0 +8 \log M\rho +\Phi_1]
\label{in1}
\end{equation}
where p. th. refers to perturbation theory, $|0>_T$ is the vacuum
after time T, $d^4x$ is the measure of integration over the four
coordinates of the centre of the instanton, $\rho$ is the scale of the
instanton, $M$ is the introduced cut-off parameter and
$S_0=\frac{8\pi^2}{g^2}$.  $\Phi_1$ denotes the contribution of the
positive frequency modes.  In the limit $M\rho >>1$, $\Phi_1$ was
evaluated to the one loop level by means of ordinary perturbation
theory to be
\begin{equation}
\Phi_1=\frac{2}{3}\log M\rho
\label{in2}
\end{equation}
Substituting (\ref{in2}) into the argument of the exponential in
(\ref{in1}), the result of the renormalised charge is obtained,
\begin{equation}
\frac{8\pi^2}{g^2(\rho)}=\frac{8\pi^2}{g_0^2} -2(4-\frac{1}{3})\log
M\rho
\end{equation} 
The $4$ explicitly came from the evaluation of the zero frequency modes
and the $-\frac{1}{3}$ from the evaluation of the transverse positive
frequency modes.

\newsection{Conclusion}

The variational ansatz of \cite{4} has been studied and the
renormalisation of the effective charge in the wavefunctional has been
carried out up to $O(g^2)$.  In this procedure, the
effective action
is a non-local, non-linear sigma model in three dimensions where the
fields considered are the group elements of the original gauge transformation.
The group elements are decomposed into low and high momentum dependent
components and the renormalisation transformation is effected by
integrating over the high momentum dependent modes up to (quadratic)
terms of $O(g^2)$.  The $\beta$
function is found to be 
\begin{equation}
\beta(g)=-\frac{g^3}{(4\pi)^2}4C_2(G) +O(g^5)
\end{equation}
This should be directly compared with the known $\beta$ function for
QCD, \cite{8};
\begin{equation}
\beta(g)=-\frac{g^3}{(4\pi)^2}[(4-\frac{1}{3})C_2(G)-\frac{2}{3}n_fC(r)]+O(g^5)
\end{equation}
The only difference between the two (in the absence of fermions
$n_f=0$) is the inclusion of the factor
$\frac{1}{3}$ in the latter, which is due to the screening effect of
virtual tranverse gluons.  Only longitudinal gluons were incorporated
in the ansatz as they satisfy Gauss' law, the constraint ensuring
gauge invariance of the wavefunctional.

The interpretation of the renormalisation procedure by
considering the Feynman diagrams shows that a vital contribution is
made by tadpole diagrams and new `horse-shoe' diagrams
which must be interpreted with the non-local propagator $G^{-1}(x-y)$.
The renormalistion scale is found to be $M$ when $|x-y|>\frac{1}{M}$
and $\frac{1}{|x-y|}$ when $|x-y|<\frac{1}{M}$, where $M$ is a
dynamically generated mass scale (and we have taken $\mu$, a
dimensionless constant from the previous section, to be 1 for simplicity).
 The explicit calculations made in this paper give a proof of the
conjecture made in \cite{4} that to calculate the gap in a variational
 approach one can use an effective QCD coupling constant and that this
 gap will automatically be related to $\Lambda_{QCD}$. 

{\bf Acknowledgements.}
 One of us (I.K) would like to thank A. Kovner for interesting
 and stimulating  discussions  about this and related subjects.  The
other (W.E.B.) wishes to thank P.P.A.R.C. for a research studentship.
\noindent

\bigskip

{\renewcommand{\Large}{\normalsize}

\end{document}